\documentclass[twocolumn,english,prl]{revtex4}
\usepackage[T1]{fontenc}
\usepackage[latin1]{inputenc}
\usepackage{amsmath}
\usepackage{graphicx}
\usepackage{amssymb}

\makeatletter
\usepackage{graphicx}
\usepackage{graphicx}

\usepackage{babel}
\makeatother
\begin{document}

\title{Coherent Destruction of Photon Emission from a Single Molecule Source}

\author{Igor Rozhkov, E. Barkai}

\affiliation{Department of Chemistry and Biochemistry, Notre Dame University,
Notre Dame, IN 46556}

\begin{abstract}
The behavior of a single molecule driven simultaneously by a laser
and by an electric radio frequency field is investigated using a non-Hermitian
Hamiltonian approach. Employing the renormalization group method for
differential equations we calculate the average waiting time for the
first photon emission event to occur, and determine the conditions
for the suppression and enhancement of photon emission. An abrupt
transition from localization-like behavior to delocalization behavior
is found.
\end{abstract}
\maketitle
Periodic driving of a two level system can significantly alter its
dynamics. \textbf{}Two renowned examples are: dynamical localization
of ultra-cold atoms in magneto-optical traps \cite{Stockmann}, and
coherent destruction of tunneling (CDT) for a particle (e.g. an electron)
in a bistable system \cite{Grossmann1}. In the latter example, a
sinusoidal external driving field localizes the electron in one of
the wells of a double well potential. This is done via particular
choice of parameters of the driving field \cite{Grossmann1}. \textbf{}Such
a coherent localization behavior was recently addressed in Ref. \cite{MakarovandMetiu}
in the context of single molecule spectroscopy \cite{OrritReview}.

When a molecule interacts with a continuous wave laser field and a
much slower radio frequency (rf) field, one can control the emission
of the molecule by the appropriate selection of the rf field parameters
(see \cite{MakarovandMetiu}). \textbf{}Coherent localization in this
case manifests itself as the divergence of average waiting time for
a single photon emission, even when the laser frequency is in resonance
with the electronic absorption frequency of the molecule. \textbf{}The
rf field effectively localizes the wave function of the molecule in
its ground state (i.e. for certain special parameters of the rf driving
field). Note that control of the emission by single molecules using
rf fields has gained recent interest since it can be used for single
photon control and hence, possibly, for quantum cryptology and quantum
computing \cite{OrritReview,OrritSph}.

In this Letter we present the criterion for the destruction and enhancement
of emission from a single molecule source interacting with an on-resonance
continuous wave laser and an rf field. \textbf{}The situation is different
from the CDT because the process of photon emission is dissipative
and, therefore, causes decoherence in the single molecule system.
\textbf{}Using the renormalization group (RG) method for differential
equations \cite{ChenGoldenfeldOono}, we obtain analytical expressions
for the average waiting time $\langle\tau\rangle$ of photon emission.
\textbf{}Although, the decoherence due to finite life time $1/\Gamma$
of the excited state of the molecule destroys complete localization,
in other words, $\langle\tau\rangle$ is always finite, we find that
$\langle\tau\rangle$ exhibits resonance behavior as a function of
control parameter. \textbf{}The most surprising result is the existence
of a critical value of $\Gamma/\omega_{rf}$ above which the resonance
peaks (i.e. tendency to localize) disappear. It turns out, that the
transition from coherent localization-like behavior to noncoherent
and delocalized behavior is not smooth\textbf{. }

Our description of a single molecule system is based on the quantum
jump approach \cite{QuantumJump}. For a molecule with a ground electronic
state $\left|g\right\rangle $ and an excited electronic state $\left|e\right\rangle $,
the Schrödinger equation is\begin{align}
i\frac{\partial}{\partial t}\left|\Psi\left(t\right)\right\rangle = & \left\{ V_{g}\cos\omega_{rf}t\left|g\right\rangle \left\langle g\right|+\frac{\Omega}{2}\left(\left|g\right\rangle \left\langle e\right|+\left|e\right\rangle \left\langle g\right|\right)\right.\nonumber \\
+ & \left.\left(V_{e}\cos\omega_{rf}t-i\frac{\Gamma}{2}+\delta\right)\left|e\right\rangle \left\langle e\right|\right\} \left|\Psi\left(t\right)\right\rangle .\label{SrodingerEqn}\end{align}
 Here $\delta$ is the laser detuning, $V_{e,g}=\boldsymbol\mu_{e,g}\cdot\boldsymbol E_{rf}$,
$\Omega=\boldsymbol\mu_{eg}\cdot\mathbf{E}$ (the Rabi frequency),
where $\boldsymbol\mu_{e,g}$ are permanent dipole moments of the
molecule in states $\left|e\right\rangle $ and $\left|g\right\rangle $,
and $\boldsymbol\mu_{eg}$ is the transition dipole, $\boldsymbol E$
and $\boldsymbol E_{rf}$ stand for the amplitudes of the laser and
the rf fields respectively. The Hamiltonian in Eq. (\ref{SrodingerEqn})
is the non-Hermitian quantum jump Hamiltonian, used in Monte Carlo
simulations of Ref. \cite{MakarovandMetiu} to propagate the {}``wave
function'' $\left|\Psi\left(t\right)\right\rangle =\Psi_{g}\left(t\right)\left|g\right\rangle +\Psi_{e}\left(t\right)\left|e\right\rangle $
\cite{QuantumJump}. In Eq. (\ref{SrodingerEqn}) we used the rotating
wave approximation for the fast on-resonant laser field, but not for
the slow off-resonant rf field. \textbf{} Although, the non-Hermitian
Hamiltonian in Eq. (\ref{SrodingerEqn}) seems simplistic, it describes
single molecule experiment very well as we demonstrate below.

The chief quantity of interest is the survival probability $P_{0}\left(t\right)=\left\langle \Psi\left(t\right)\right|\left|\Psi\left(t\right)\right\rangle $,
i.e. the probability of no emission event to occur in the time interval
between $0$ and $t$. The knowledge of $P_{0}\left(t\right)$ enables
us to compute statistics of quantum jumps from state $\left|e\right\rangle $
to state $\left|g\right\rangle $, i.e. the average waiting time for
emission event \cite{MakarovandMetiu,QuantumJump}\begin{equation}
\left\langle \tau\right\rangle =\int_{0}^{\infty}\left(\left|\Psi_{g}\left(t\right)\right|^{2}+\left|\Psi_{e}\left(t\right)\right|^{2}\right)dt.\label{tau1}\end{equation}
 In what follows\textbf{,} we first consider the case of $\Omega\ll\Gamma,\omega_{rf}$,
and later, the case of $\Gamma<\Omega\ll\omega_{rf}$. We solve the
Schrödinger equation (\ref{SrodingerEqn}) perturbatively employing
the renormalization group method \cite{ChenGoldenfeldOono} (cf. recent
solution for CDT problem \cite{Grossmann1} with the RG procedure
\cite{Frasca2}). Note that the limit of small Rabi frequency ($\Omega\ll\omega_{rf}$)
corresponds to the limit of high frequency driving field considered
previously in the CDT problem \cite{Grossmann1}. \textbf{}We study
the effect of the non-Hermitian part of the Hamiltonian on localization
behavior in the same limit.

Introducing scalings $t\mapsto\omega_{rf}t$, $V_{e,g}\mapsto V_{e,g}/\omega_{rf}$,
$\delta\mapsto\delta/\omega_{rf}$, $\Gamma\mapsto\Gamma/\omega_{rf}$,$\Omega\mapsto\epsilon\Omega/\omega_{rf}$
(where $\epsilon\ll1$) we pass to the {}``amplitude''-variables:
$\Psi_{g}\left(t\right)=c_{g}\left(t\right)\textrm{exp}\left\{ -iV_{g}\sin t\right\} $,
$\Psi_{e}\left(t\right)=c_{e}\left(t\right)\textrm{exp}\left\{ -iV_{e}\sin t-\Gamma t/2\right\} $.
Thus, for the case of zero detuning $\delta=0$, we arrive at the
set of coupled ordinary differential equations:\[
\frac{d}{dt}c_{g}=-i\epsilon c_{e}\frac{\Omega}{2}\sum_{k=-\infty}^{\infty}\textrm{e}^{-ikt-\Gamma t/2}J_{k}\left(\xi\right),\]
\begin{equation}
\frac{d}{dt}c_{e}{\displaystyle =-i\epsilon c_{g}\frac{\Omega}{2}\sum_{k=-\infty}^{\infty}\textrm{e}^{ikt+\Gamma t/2}J_{k}\left(\xi\right)}.\label{systemEq2}\end{equation}
 where we expanded exponents in series of Bessel functions of the
first kind $J_{k}$ \cite{Abramowitz}. The modulation index $\xi=V_{e}-V_{g}$
is a parameter governing the amplitude of the driving rf field.

To solve Eqs. (\ref{systemEq2}) we proceed with the application of
the RG approach \cite{ChenGoldenfeldOono}. The naive solution to
Eqs. (\ref{systemEq2}) is given by a power series in small $\epsilon$:
$C\left(t\right)=\left(c_{g},c_{e}\right)=c^{\left(0\right)}\left(t\right)+\epsilon c^{\left(1\right)}\left(t\right)+O\left(\epsilon^{2}\right)$,
provided that $c^{\left(n\right)}\left(t\right)$ are $O\left(1\right)$
for all $t$. At large enough times, however, one or several $c^{\left(n\right)}\left(t\right)$
($n>1$) may become greater than $c^{\left(0\right)}\left(t\right)$.
Such terms, responsible for the break-down of the naive expansion
are called secular terms. The RG procedure, we carry out below, regularizes
the naive expansion by identifying the secular terms and eliminating
them \cite{ChenGoldenfeldOono}. This technique enables us to obtain
the long time asymptotic solution. 

At $O\left(\epsilon^{0}\right)$, the solution of Eq. (\ref{systemEq2})
is simply a constant vector $c^{\left(0\right)}=\left(A,B\right)^{T}$.
Substituting this result into the next order equations, and integrating,
we have \[
c_{g}^{\left(1\right)}=-i\Omega B\sum_{k=-\infty}^{\infty}J_{k}\left(\xi\right)\left(1-\textrm{e}^{-\Gamma t/2-ikt}\right)\frac{\Gamma-2ik}{\Gamma^{2}+4k^{2}},\]
 \begin{equation}
c_{e}^{\left(1\right)}=-i\Omega A\sum_{k=-\infty}^{\infty}J_{k}\left(\xi\right)\left(\textrm{e}^{\Gamma t/2+ikt}-1\right)\frac{\Gamma-2ik}{\Gamma^{2}+4k^{2}}.\label{CEG1}\end{equation}
 Further application of the naive procedure produces terms, which
grow with time $t$, i.e. secular terms\[
c_{g}^{\left(2\right)}=-\frac{t\Omega^{2}}{2}A\sum_{k=-\infty}^{\infty}J_{k}^{2}\left(\xi\right)\frac{\Gamma}{\Gamma^{2}+4k^{2}}+NST,\]
 \begin{equation}
c_{e}^{\left(2\right)}=-\frac{t\Omega^{2}}{2}B\sum_{k=-\infty}^{\infty}J_{k}^{2}\left(\xi\right)\frac{\Gamma}{\Gamma^{2}+4k^{2}}+NST,\label{CEG2}\end{equation}
 where \emph{NST} stands for nonsecular terms. The following RG equations
for variables $A$ and $B$ are intended to regularize the naive expansion
(to order $O\left(\epsilon^{3}\right)$) \cite{ChenGoldenfeldOono}:\[
\frac{dA_{R}}{dt}=-\frac{\epsilon^{2}\Omega^{2}\Gamma}{2}A_{R}\sum_{k=-\infty}^{\infty}\frac{J_{k}^{2}\left(\xi\right)}{\Gamma^{2}+4k^{2}},\]
 \[
\frac{dB_{R}}{dt}=-\frac{\epsilon^{2}\Omega^{2}\Gamma}{2}B_{R}\sum_{k=-\infty}^{\infty}\frac{J_{k}^{2}\left(\xi\right)}{\Gamma^{2}+4k^{2}},\]
 where subscript {}``\emph{R}'' indicates, that we are solving for
the renormalized values of $A$ and $B$. Thus, to $O\left(\epsilon^{3}\right)$
$A_{R}$ and $B_{R}$ satisfy\[
A_{R}\left(t\right)=A\left(0\right)\exp\left\{ -\frac{\epsilon^{2}\Omega^{2}\Gamma t}{2}\sum_{k=-\infty}^{\infty}\frac{J_{k}^{2}\left(\xi\right)}{\Gamma^{2}+4k^{2}}\right\} ,\]
 \begin{equation}
B_{R}\left(t\right)=B\left(0\right)\exp\left\{ -\frac{\epsilon^{2}\Omega^{2}\Gamma t}{2}\sum_{k=-\infty}^{\infty}\frac{J_{k}^{2}\left(\xi\right)}{\Gamma^{2}+4k^{2}}\right\} .\label{AandB}\end{equation}
 At time $t=0$ we assume that the molecule is in the ground state,
hence $c_{e}\left(0\right)=0,c_{g}\left(0\right)=1$. Applying these
initial conditions and replacing $A$ and $B$ with their renormalized
values (Eqs. (\ref{AandB})) in $c^{\left(0\right)}$ and in Eq. (\ref{CEG1})
we obtain $O\left(\epsilon^{2}\right)$ perturbation results for amplitudes
$c_{e}$ and $c_{g}$. Then, we switch back to the original variables
$\Psi_{e,g}\left(t\right)$ and arrive at \[
\Psi_{g}\left(t\right)=\textrm{e}^{-iV_{g}\sin t-\zeta t}+O\left(\epsilon^{2}\right),\]
 \[
\zeta=\frac{\epsilon^{2}\Omega^{2}}{2}\sum_{k=-\infty}^{\infty}J_{k}^{2}\left(\xi\right)\frac{\Gamma}{\Gamma^{2}+4k^{2}},\]
\begin{align}
\Psi_{e}\left(t\right) & =-i\epsilon\Omega\textrm{e}^{-iV_{e}\sin t-\Gamma t/2-\zeta t}\sum_{k=-\infty}^{\infty}J_{k}\left(\xi\right)\nonumber \\
 & \times\frac{\left(\textrm{e}^{\Gamma t/2+ikt}-1\right)\left(\Gamma-2ik\right)}{\Gamma^{2}+4k^{2}}+O\left(\epsilon^{3}\right).\label{Psie}\end{align}
 For convenience we set $\epsilon=1$, keeping in mind \textbf{}that
the Rabi frequency $\Omega$ is small. \textbf{}Then, using Eq. (\ref{tau1},
\ref{Psie}) we calculate the mean waiting time $\langle\tau\rangle$
\begin{figure}[htbp]
\begin{center}\includegraphics{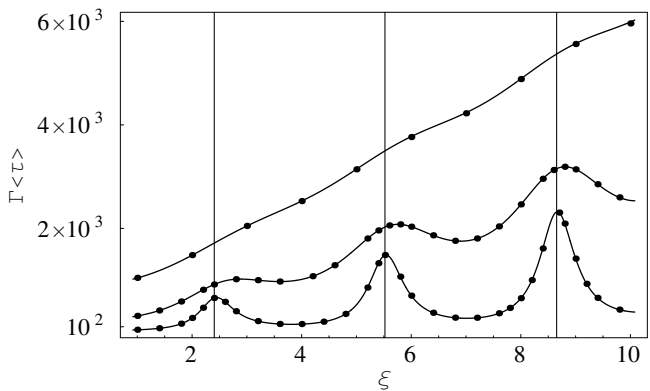}\end{center}

\caption{\label{MTau} Scaled mean time between emission $\Gamma\left\langle \tau\right\rangle $
is plotted as a function of modulation index $\xi$. Solid line represents
RG prediction of Eq. (\ref{tbar3}), dots correspond to the numerical
solution of Eq. (\ref{SrodingerEqn}). Rabi frequency $\Omega=0.1$.
Radiative decay rates are $\Gamma=0.5$; $1.5$; $3.0$ (from top
to bottom). Grid-lines indicate zeroes of $J_{0}\left(\xi\right)$.}
\end{figure}

\begin{figure}[htbp]
\begin{center}\includegraphics{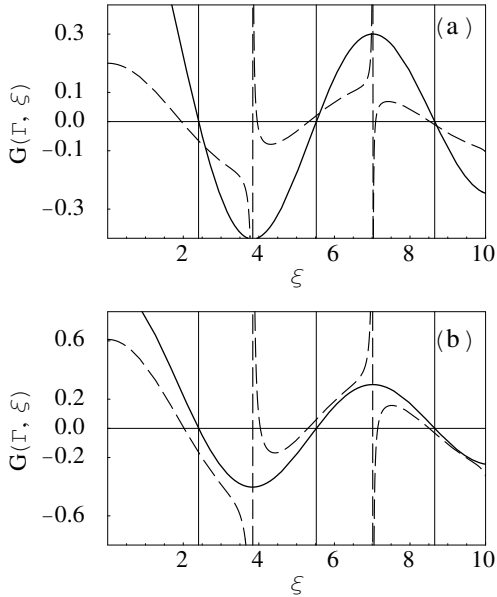}\end{center}

\caption{\label{Crossings} Illustration to Eq. (\ref{nun}). Bessel function
$J_{0}$ (solid line) plotted as a function of $\xi$ together with
the right hand side of Eq. (\ref{nun}), $G\left(\Gamma,\xi\right)$.
Radiative decay rates are (a) $\Gamma=1.0$, (b) $\Gamma=2.5$. Grid-lines
indicate zeroes of $J_{0}\left(\xi\right)$.}
\end{figure}

\begin{equation}
\left\langle \tau\right\rangle ^{-1}=\Gamma\Omega^{2}\sum_{k=-\infty}^{\infty}\frac{J_{k}^{2}\left(\xi\right)}{\Gamma^{2}+4k^{2}}+O\left(\frac{\Omega^{2}}{\Gamma^{2}}\right).\label{tbar3}\end{equation}
 In Fig. \ref{MTau} we plot the scaled mean time $\Gamma\left\langle \tau\right\rangle $,
given by Eq. (\ref{tbar3}) as a function of the modulation index
$\xi$, and compare this prediction with the results of numerical
solution of Eq. (\ref{SrodingerEqn}). We observe a good agreement
between the numerical and analytical results, indicating that the
RG approach indeed captured the global behavior of $\Psi_{e,g}$.
We also observe, that for the smallest value of the decay rate $\Gamma$
($\Gamma/\omega_{rf}$ in original units), maxima of $\left\langle \tau\right\rangle $
occur close to the zeroes of the Bessel function $J_{0}\left(\xi\right)$,
but with increase in $\Gamma$, the maxima of $\left\langle \tau\right\rangle $
shift and broaden. 

\begin{figure}[htbp]
\begin{center}\includegraphics{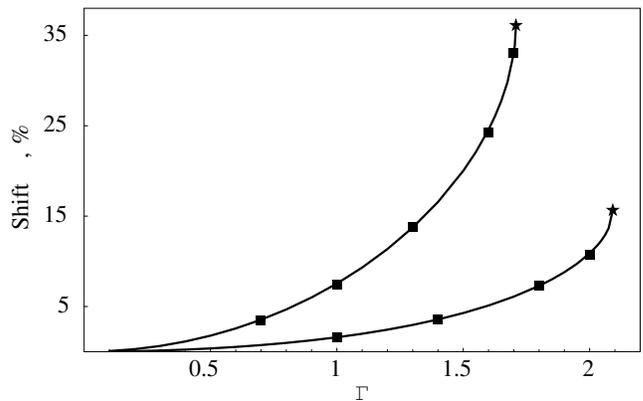}\end{center}

\caption{\label{Shifts}The shifts $\left(\xi-\xi_{n}\right)/\xi_{n}$ found
from Eq. (\ref{nun}) (solid lines) are compared to those found from
numerical solution of Eq. (\ref{SrodingerEqn}) (boxes). They are
plotted as a function of $\Gamma$ ($\Gamma/\omega_{rf}$ in original
units) for the peaks close to the first (upper curve) and the second
(lower curve) zeroes of $J_{0}$ (the first two peaks in Fig. \ref{MTau}).
Critical points are marked by stars.}
\end{figure}

To determine the positions of the extremums of $\left\langle \tau\right\rangle $
as a function of $\xi$ (Fig. \ref{MTau}) and to investigate the
effect of the dissipation on the emission suppression (enhancement)
criterion, we use Eq. (\ref{tbar3}) and set $d\left\langle \tau\right\rangle /d\xi=0$.
This condition results in the following transcendental equation,\begin{align}
J_{0}\left(\xi\right) & =\frac{\Gamma^{2}}{J_{1}\left(\xi\right)}\sum_{k=1}^{\infty}\frac{J_{k}\left(\xi\right)\left(J_{k-1}\left(\xi\right)-J_{k+1}\left(\xi\right)\right)}{\Gamma^{2}+4k^{2}}\nonumber \\
 & =G\left(\Gamma,\xi\right).\label{nun}\end{align}
 Solution of Eq. (\ref{nun}) yields values of $\xi$ which minimize
and maximize $\langle\tau\rangle$. These values correspond to enhancement
and reduction of photon emission rate $1/\langle\tau\rangle$.

In the limit of $\Gamma\rightarrow0$ we recover the well-known condition,
obtained previously in the context of CDT\textbf{,} $J_{0}\left(\xi_{n}\right)=0$
($\xi_{n}$ are the zeros of the Bessel function $J_{0}$). For small
$\Gamma$, since shifts in $\xi$ with respect to $\xi_{n}$ are small,
we can expand both sides of Eq. (\ref{nun}) in Taylor series around
points $\xi_{n}$. Then, the shifts of maxima are given by\begin{equation}
\xi-\xi_{n}\sim\frac{\Gamma^{2}}{J_{1}^{2}\left(\xi_{n}\right)}\sum_{k=1}^{\infty}\frac{J_{k}\left(\xi_{n}\right)\left(J_{k+1}\left(\xi_{n}\right)-J_{k-1}\left(\xi_{n}\right)\right)}{\Gamma^{2}+4k^{2}}.\label{smallGammashifts}\end{equation}
For arbitrary values of $\Gamma$, Eq. (\ref{nun}) has to be solved
numerically. In Fig. \ref{Crossings} we illustrate this solution
by plotting $G\left(\Gamma,\xi\right)$ and $J_{0}\left(\xi\right)$
versus $\xi$ for two values of $\Gamma$. In Fig. \ref{Crossings}a
($\Gamma=1.0$) we observe the graphs crossings close to the roots
of Bessel function $J_{0}$. However, as we increase $\Gamma$ to
$2.5$, crossings disappear (Fig. \ref{Crossings}b). In other words,
Eq. (\ref{nun}) does not have solutions for this value of $\Gamma$,
and first two maxima of $\left\langle \tau\right\rangle $ do not
exist. A similar transition also happens with the minima of $\left\langle \tau\right\rangle $.
As it follows from Eq. (\ref{nun}), the minima are found in the vicinity
of zeroes of $J_{1}$ for $\Gamma$ close to zero, and disappear for
sufficiently large $\Gamma$ (see Fig. \ref{Crossings}b).

This outcome is an indication of unexpected critical behavior of $\left\langle \tau\right\rangle $
in $\Gamma$. Maxima (minima) of the waiting time $\langle\tau\rangle$
do not disappear gradually with the increasing decay rate, due to
the peaks broadening. Instead, there exist a critical value of $\Gamma$
for each extremum, which is equivalent to existence of infinite number
of critical points\textbf{.} We note, that in the vicinity of critical
point $\left(\xi_{cr},\Gamma_{cr}\right)$, the straight forward expansion
of Eq. (\ref{nun}) yields $\xi_{cr}-\xi\propto\left(\Gamma_{cr}-\Gamma\right)^{\beta}$,
with $\beta=1/2$. Furthermore, as can be concluded from Fig. \ref{Crossings},
the neighboring pairs of minima and maxima disappear at the same values
of $\Gamma_{cr}$, that is, maxima and minima flow towards each other
and annihilate. In Fig. \ref{Shifts} we plot shifts in the positions
of maxima with respect to $\xi_{n}$ given by numerical solution of
Eq. (\ref{nun}) as a function of $\Gamma$. Both curves in Fig. \ref{Shifts}
display excellent agreement with the results of numerical solution
of the Schrödinger equation (\ref{SrodingerEqn}).

We remark, that the RG procedure for the case of nonzero detuning
is performed in a similar fashion and results in\begin{equation}
\left\langle \tau\right\rangle ^{-1}=\Gamma\Omega^{2}\sum_{k=-\infty}^{\infty}\frac{J_{k}^{2}\left(\xi\right)}{\Gamma^{2}+4\left(k-\delta\right)^{2}}+O\left(\frac{\Omega^{2}}{\Gamma^{2}}\right).\label{flspectrabigGm}\end{equation}
 In this case, maxima of mean waiting time occur close to zeroes of
$J_{\delta}\left(\xi\right)$ ($\delta$ is an integer), if $\Gamma$
is not large. Otherwise, they shift according to the $\delta=0$ scenario.

So far we have considered the situation when Rabi excitation frequency
$\Omega$ is smaller than $\Gamma$. In a view of single molecule
experiments carried out in Ref. \cite{Orrit}, an opposite situation
$\Omega>\Gamma$ is also important. Assuming, that $\Omega,\Gamma\ll\omega_{rf}$
we now turn to the criterion for the destruction of emission for this
case, and verify the compatibility of the RG results and of our model
Hamiltonian with the available experimental data.

Note, that if $\Gamma$ is smaller than $\omega_{rf}$, we need to
modify our derivation. This time, we keep the laser detuning $\delta$
in the \emph{ee}-matrix element of the Hamiltonian in Eq. (\ref{SrodingerEqn}),
and switch to {}``amplitudes'': $\Psi_{g}\left(t\right)=c_{g}\left(t\right)\textrm{exp}\left\{ -iV_{g}\sin\omega_{rf}t\right\} $,
$\Psi_{e}\left(t\right)=c_{e}\left(t\right)\textrm{exp}\left\{ -iV_{e}\sin\omega_{rf}t-\Gamma t/2-i\delta t\right\} $.
We use different scaling to ensure that the rf field frequency is
much greater than both $\Omega$ and $\Gamma$ (i.e. $\Gamma\mapsto\epsilon\Gamma/\omega_{rf}$,
$\Omega\mapsto\epsilon\Omega/\omega_{rf}$). Our system of ordinary
differential equations now reads\[
\frac{dc_{g}\left(t\right)}{dt}=-\frac{i\epsilon\Omega}{2}{\displaystyle \sum_{k=-\infty}^{\infty}}\textrm{e}^{-ikt-i\delta t}J_{k}\left(\xi\right)c_{e}\left(t\right),\]
\[
\frac{dc_{e}\left(t\right)}{dt}=-\frac{i\epsilon\Omega}{2}{\displaystyle \sum_{k=-\infty}^{\infty}}\textrm{e}^{ikt+i\delta t}J_{k}\left(\xi\right)c_{g}\left(t\right)-\frac{\epsilon\Gamma}{2}c_{e}\left(t\right),\]
where all the quantities are in units of $\omega_{rf}$. Unlike the
previous case (Eqs. (\ref{CEG1}, \ref{CEG2})), the secular terms
appear (at integer values of $\delta$) in the naive expansion already
at the first order. The details of calculations with the RG method
will be given elsewhere, here we only present the final expression
for the average waiting time:%
\begin{figure}[htbp]
\begin{center}\includegraphics{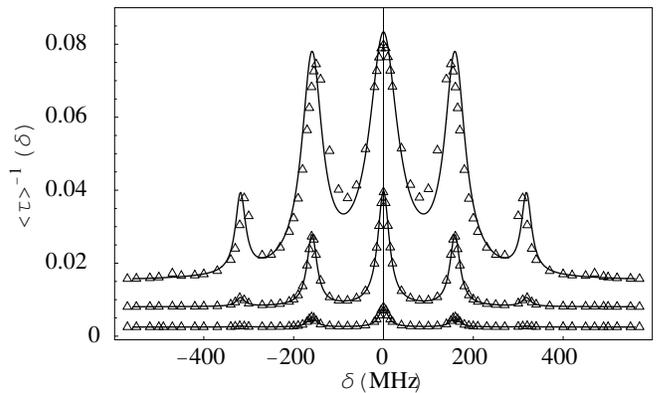}\end{center}

\caption{\label{OrritCase} Comparison of the predictions for $\left\langle \tau\right\rangle ^{-1}\left(\delta\right)$
given by Eq. (\ref{fluorescence1}) (solid lines) with the excited
state population calculated from Bloch equations (Ref. \cite{Orrit})
for various values of Rabi frequencies (from bottom to the top): $\Omega=0.29;0.9;3.2$
in units of $\Gamma$. Parameters are taken from experiment of Ref.
\cite{Orrit}: $\omega_{rf}/2\pi=140$ MHz, $\xi=1.14$, $\Gamma/2\pi=20$
MHz (empty triangles). The results are arbitrarily shifted vertically
for transparency, therefore, units on vertical axis are arbitrary.
Our predictions are in accord with experiments \cite{Orrit}.}
\end{figure}
 \begin{equation}
\left\langle \tau\right\rangle ^{-1}\left(\delta\right)\approx{\displaystyle \sum_{k}}\frac{\Gamma\Omega^{2}J_{k}\left(\xi\right)^{2}}{\Gamma^{2}+2\Omega^{2}J_{k}\left(\xi\right)^{2}+4\left(k-\delta\right)^{2}}.\label{fluorescence1}\end{equation}
 The approximate equality sign in Eq. (\ref{fluorescence1}) indicates,
that it is only valid when resonances are not overlapping, which is
consistent with our assumption $\Gamma\ll\omega_{rf}$. We see that
the criterion for destruction of emission is given by $J_{k}\left(\xi\right)=0$
for integer values of the detuning $\delta=k$ (in units of $\omega_{rf}$).

Now we are ready to compare our theory with experiment of Ref. \cite{Orrit}.
The experimental results had been previously matched with the numerical
solution of optical Bloch equations \cite{Orrit}. We repeat the numerical
simulations carried out in Ref. \cite{Orrit}, and compare the results
with analytical prediction of Eq. (\ref{fluorescence1}) in Fig. \ref{OrritCase}
\cite{footnote}. All three graphs in Fig. \ref{OrritCase} illustrate
nearly perfect agreement between the theory and experiment, suggesting
that the observation of critical behavior addressed in this Letter
may be possible.

This work was supported by National Science Foundation award CHE-0344930.


\begin{thebibliography}{10}
\bibitem{footnote}We also compared numerical solution of Eq. (\ref{SrodingerEqn}) with
RG solution, and found a good agreement.
\end{thebibliography}
\end{document}